\documentclass[11pt]{article}
\usepackage{amssymb,setspace}

  \makeatletter
    \def\thebibliography#1{\section*{REFERENCES\@mkboth
      {REFERENCES}{REFERENCES}}\list
      {[\arabic{enumi}]}{\settowidth\labelwidth{[#1]}\leftmargin\labelwidth
        \advance\leftmargin\labelsep
        \usecounter{enumi}}
        \def\newblock{\hskip .11em plus .33em minus .07em}
        \sloppy\clubpenalty4000\widowpenalty4000
        \sfcode`\.=1000\relax}
  \makeatother
\begin{document}

\bibliographystyle{unsrt} 

\title{A spatially-VSL gravity model\\ with 1-PN limit of GRT}
\author{Jan Broekaert\footnote{CLEA-FUND, Vrije Universiteit Brussel.  }}
\maketitle

\begin{abstract}
\noindent
In the \emph{static} field configuration, a  spatially-Variable Speed of Light (VSL) scalar gravity  model  with Lorentz-Poincar\'e interpretation   was shown to  reproduce the phenomenology implied by the Schwarzschild  metric.   In the present development, we effectively  cover   configurations  with source kinematics  due to an induced \emph{sweep velocity} field  $\mathbf{w}$. The scalar-vector model now provides a Hamiltonian description for particles  and  photons in full accordance with the first Post-Newtonian  (1-PN)  approximation of  General Relativity Theory (GRT).   This result requires the validity of Poincar\'e's Principle of Relativity, \emph{i.e.} the unobservability  of `preferred' frame movement. Poincar\'e's principle fixes the amplitude of the sweep velocity field of the moving source, or equivalently  the `vector potential' $ \mbox{\boldmath$\xi$}$ of GRT (e.g.;  S. Weinberg, \emph{Gravitation and cosmology}, 1972), and   provides the correct 1-PN limit of GRT.   The  implementation of this principle requires  acceleration transformations derived from gravitationally modified Lorentz transformations. A comparison with the acceleration transformation  in GRT  is done.  The  present scope of the model is  limited to weak-field  gravitation without retardation and with gravitating test bodies. In conclusion the model's merits in terms of  a simpler space, time and gravitation ontology ---in terms of a Lorentz-Poincar\'e--type interpretation--- are explained  (e.g.  for `frame dragging', `harmonic coordinate condition').

%\PACS{04.20.-q, 04.50.+h}

%\keywords{modified Lorentz Transformations, scalar--vector gravitation}
\end{abstract}

\section{Introduction} 
Modeling relativistic gravitation  in  flat, ``unrenormalized", 
space-time has a long tradition starting with pre-general relativistic  approaches  of Lorentz,  Poincar\'e, Einstein, Abraham  and Nordstr\"om \cite{prerelativisticapproaches} and finalized in the field-theoretic approach of Fierz, Rosen, Gupta and many, more recent, authors  \cite{fieldapproachesgrouped}  (see references in  Cavalleri and Spinelli's work \cite{CavalleriSpinelli1980}).   In these latter theories the \emph{physically observed} space-time ---using ``measuring rods and clocks" ---  still turns out to be Riemannian. This is because the measurements are modified by the ``universal" action of gravitation, which leads to the ``renormalized" curved space.  It is well known that a \emph{second rank} tensor theory is required to describe all current gravitational experiments adequately and, have convergence with Einstein's theory (e.g. \cite{CavalleriSpinelli1980}).  We have shown previously  however that a 0-rank (scalar) gravitation model, with spatially-variable speed of light and gravitationally modified Lorentz transformation  is still able to  reproduce the four `basic' experiments of GRT \cite{Broekaert2002, Broekaert2005a}. At present we will expand the model to encompass the full first Post-Newtonian  dynamics of GRT.\\
We briefly recall that one of the basic premisses of the model is Poincar\'e's \emph{conventionalism of geometry} \cite{conventionalismgrouped}. This principle  purports the  compensation of the adopted basic geometry by a conjugated gravitational dynamics and congruence relation \cite{Poincare1905, Einstein1921, Gruenbaum1973} while retaining empirical indistinguishability with GRT.  A second premiss is to express the congruence relation explicitly by maintaining the Lorentz-Poincar\'e interpretation of SRT in GRT.  \emph{I.e.} we render explicit the gravitational affecting of physical space and time observations ---the contraction and dilation of  ``measuring rods and clocks". This leads to the use of two types of gravitationally modified Lorentz transformations,  relating both;  the  affected (or normalized) and unaffected (or unrenormalized) perspective and, by composition of the previous, non-identical observers in the affected perspective. We emphasize that naturally the unaffected perspective cannot 
be realized by any \emph{physical} observer, it merely corresponds to the coordinate space description. Finally, these modifications are due to both; the  incorporated  spatially variable speed of light and, position and time dependent scaling functions. While this latter feature is similar to  some precursor  theories  of gravitation \cite{prerelativisticapproaches}, our Lorentz-Poincar\'e type (``L-P") model still maintains  the invariance of the locally observed velocity of light  and local Lorentz symmetry. \\
The term ``variable speed of light" is a historic concept already  occurring in the precursory work of  Einstein \cite{Einstein1907} in 1907 where  the velocity of light  depends on the gravitational field (or potential).  The term was used \emph{in tempore non suspecto} by Wei-Tou Ni \cite{Ni1972} in order to classify a number of alternative gravitational theories. The same term ---now written by its acronym--- ``VSL" is used in some recent  work where it is understood as an \emph{epochal} phenomenon; i.e. exposing dependence on a  cosmological time scale  \cite{Moffat1993, AlbrechtMagueijo1999,Magueijo2003}. In the present L-P model  however we take \emph{ab initio} what in GRT is found to be the case in photon dynamics; namely that the velocity of light ---or a photon on its geodesic--- in coordinate space is not constant, but that it is depending on the gravitational field, implying dependence on  its location and instant  in coordinate space and time. Excerpts revealing this interpretation of spatial-VSL in mainstream GRT works are given in the bibliography \cite{VSL2004}.\\
The main aim of the present work is to extend the scalar model which  previously only covered the static field configuration (short recapitulation of  \cite{Broekaert2005a} in Section \ref{Newtonfit}). Now configurations with proper source kinematics are included, e.g. Earth rotation.
Our model thus assigns to a moving source a scalar and a vector  potential $\{\varphi, \mathbf{w}\}$. The vector potential will describe  the `effective' velocity of a test particle in the gravitational field due to source movement. The source's resultant \emph{sweep velocity} field expresses the test particle's   source-relative movement in contrast to   `preferred'  frame-relative movement (Section \ref{sweep}).  A Hamiltonian expression for particles and photons till first order in the sweep velocity  is obtained by considering ---relative to the unaffected observer--- a Galilean composition of the energy and momentum.  The equations of motion result in a dynamical acceleration which formally is the first Post-Newtonian expression of the acceleration in GRT (Subsection \ref{PN}).  Subsequently this correspondence is rendered numerically exact by assuming the validity of the  Poincar\'e Principle of  Relativity (PPR) (Section \ref{PPR}).  Its implementation requires an adequate interpretation of the acceleration transformation which will be compared to its equivalent in  GRT (Section \ref{ACCELERATION}).  Finally some examples of simpler ontological interpretations in line with the Lorentz-Poincar\'e interpretation in gravitation  are discussed in the conclusion (Section \ref{conclusion}). \\
Many alternative approaches to GRT have been formulated in the literature, some of which  share  aspects  with our approach, e.g.  interaction through gravitational potential in variable speed of light initiated by Einstein and Abraham \cite{prerelativisticapproaches},  polarizable vacuum models for gravitation starting  with Wilson \cite{Wilson1921} and Dicke \cite{Dicke1957,BransDicke1961} and more recently in Lightman and Lee \cite{LightmanLee1973} and Puthoff  \cite{Puthoff2002}, \emph{physical} substrate interpretation  of analogue gravity by  Visser {\emph{et. al.} \cite{BarceloLiberatiVisser2005}, \emph{scalar} field gravitation models  Coleman, Sj\"odin \cite{Coleman1971,Sjodin1990} and Arminjon, Evans and Nandi \cite{Arminjon2004a,scalarfieldmodelsgrouped}, \emph{nonsingular} ether gravitation models  by Cavalleri and Tonni \cite{CavalleriTonni1997} and,  \emph{field} conceptions of gravitation, e.g.  Weinberg  and others \cite{Weinberg1965,CavalleriSpinelli1980,PittsSchieve2004}.

\section{Gravitationally modified Lorentz transformations \label{GMLT}}
The congruence relations, which express the gravitational and kinematical affecting of space and time observations,  have been formulated as gravitationally modified Lorentz Transformations (GMLT). This is done by  applying  isotropic scaling functions ${\Phi\left({\bf x} \right)}^n$, $0\leq{\Phi\left({\bf x} \right)}\leq1$, to the considered physical quantities ($n$-specific), and  also to the velocity of  light ($n=2$) \cite{Broekaert2002}.  The \emph{first type} of these transformations relate quantities observed by gravitationally \emph{affected} (natural, ``renormalized" or ``local" geometry) and \emph{unaffected}  (coordinate, ``unrenormalized" or ``Newtonian" geometry)  ``observers" and, these transformation explicitly expose a spatially variable speed of light (\ref{SpaceTimeGMLTsotosac}, \ref{EnergyMomentumGMLT}). The \emph{second type} transformations shunts the unaffected perspective and relates affected observers, recovering i) the  invariance of the locally observed velocity of light,  and ii) the local Minkowski metric (appendix in \cite{Broekaert2005a}).\\
For a space and a time  interval, observed by the gravitationally unaffected {$S_0$} observer, and the same infinitesimal  space and  time interval as observed by gravitationally
affected observer {$S'$} we have set:
\begin{eqnarray} 
 \delta {\bf x} = \delta {\bf x}' {\Phi   \left( {\bf x} \right)} \ {\rm (\emph{contraction})} \ &,& \  \delta  t  = \frac{\delta t'}{ {\Phi   \left( {\bf x} \right)}} \ {\rm (\emph{dilation})} \label{congruencerelation}
\end{eqnarray}
These interval relations are related to a constrained differential of the diffeomorphism $x^\mu = x^\mu(x')$ that transforms  the coordinate space of $S'$ into the space of  $S_0$.  The contraction of rulers and the dilation of the periods of clocks are \emph{local} gravitational effects; their space and time extension should therefore be restricted  to domains over which $\Phi ({\bf x})$ can be considered sufficiently invariable to the required accuracy, and as such the intervals have to be infintesimal. The specific \emph{scaling} nature of the gravitational effect ---as in  transformation eqs. (\ref{congruencerelation})---  and observation conditions impose constraints on the interval relation which preempts the coincidence with a complete differential relation (cfr the Appendix): 
\begin{eqnarray}
\lim_{ \delta x'_i \to 0} \left(\delta t' -  d t'{\vert_{_{dx_i =0}}}\right) = 0 \ &,&\ \lim_{\delta t', \delta x'_{j \neq i} \to 0} \left( \delta \xi'_i  - d {\xi_i}{\vert_{_{dx_{j \neq i} =0, d\tau = 0}}} \right) = 0 \end{eqnarray}
The congruence relations do not allow invariance of  the velocity of light in both the  affected $S'$ and unaffected $S_0$ perspectives.  In the \emph{physical}  perspective, of the affected observer $S'$, the invariance of the locally observed velocity of light  $c'$  is secured by imposing  specific spatial variability in the \emph{unaffected} perspective of observer $S_0$: 
\begin{eqnarray}
  {c ({\bf x})} = {c '} {\Phi   \left( {\bf x} \right)}^2 \ \  ( c',\ \ {\rm the\  observed\  constant\ velocity\ of \ light)} \label{VSL}
\end{eqnarray}
\emph{i.e.} $n=2$ as mentioned previously.\\
The gravitational congruence relation (\ref{congruencerelation}), with kinematical affecting  factor $\gamma$ and light synchronization (e.g. \cite{MansouriSexl1977}), lead to the full congruence relations in the form of gravitationally modified Lorentz Transformations  of the first type, \emph{i.e.} between affected and unaffected observers \cite{Broekaert2002, Broekaert2005a}. 
The space-time  GMLT,  {$S_0$} to  {$S'$},  with relative frame velocity ${\mathbf u}$ is given by: 
\begin{eqnarray}
\delta{\bf x}'  &=&  \left(( {\delta{\bf x}_\parallel} -{\bf u} (\mathbf{x}) \delta t)  {\gamma  \left( u, {\bf x} \right)}     +
  {\delta{\bf x}_\perp} \right) \frac{1}{{\Phi   \left( {\bf x}  \right)}}  \\
\delta t' &=& \left( \delta t - \frac{ {\bf u}(\mathbf{x}) . \delta{\bf x} }{{c ({\bf x} )}^2}  \right)  {\gamma  \left( u, {\bf x} \right)}  {\Phi   \left( {\bf x}  \right)}   \label{SpaceTimeGMLTsotosac}
\end{eqnarray}
and the inverse  GMLT, {$S'$} to {$S_0$}, is;
\begin{eqnarray}
\delta{\bf x}  &=& \left(( {\delta{\bf x}_\parallel}' -{{\bf u}'}\ \delta t')  {\gamma  \left( u' \right)}     +
  {\delta{\bf x}_\perp}' \right) {\Phi   \left( {\bf x}  \right)}  \\
 \delta t &=& \left( \delta t' - \frac{ {{\bf u}'} . \delta{\bf x}' }{c'^2}  \right) \frac{ {\gamma  \left( u' \right)}}{{\Phi   \left( {\bf x}  \right)}}  
\label{SpaceTimeGMLTsactoso}
\end{eqnarray}
with ${\bf u} = {\bf u} ({\bf x} )$   satisfying:
\begin{eqnarray}
{\bf u} ({\bf x} ) \ = \ - {{\bf u}'} {\Phi \left( {\bf x}  \right)}^2  \label{framevelocities}
\end{eqnarray}
and ${\gamma  \left( u' \right)}$ and ${\gamma  \left( u, \mathbf{x}  \right)}$, by eqs. (\ref{VSL},\ref{framevelocities}), satisfy:
\begin{eqnarray}
{\gamma  \left( u, \mathbf{x} \right)} \ \equiv \ \left( 1 - u(\mathbf{x} )^2/c(\mathbf{x} )^2 \right)^{-1/2} & = &  \left( 1 - {u'}^2/{c'}^2 \right)^{-1/2} \ \equiv \ {\gamma  \left( u' \right)} .
\end{eqnarray}
We notice thus that the values of $\gamma (u')$ and $\gamma (u, \mathbf{x})$ are equal. \\
 The velocity $\mathbf{v}' = \delta{\bf x}'/\delta{\bf t}' $ in affected and,  $\mathbf{v} = \delta{\bf x}/\delta{\bf t} $ in unaffected perspective is related according:
\begin{eqnarray}
{\mathbf v}' &=& \frac{{\mathbf v}_\parallel - {\mathbf u}( \mathbf{x}) + {\mathbf v}_\perp/\gamma (u, \mathbf{x})}{\left(1 - {\mathbf u}( \mathbf{x}).{\mathbf v}/c( \mathbf{x})^2\right) \Phi( \mathbf{x})^2}.  \label{velocityrelation}
\end{eqnarray}
From the space time GMLT  (\ref{SpaceTimeGMLTsactoso}, \ref{SpaceTimeGMLTsotosac}) the \emph{contravariant} space-time GMLT, {$S'$} to {$S_0$} for gradient operators can be derived:
\begin{eqnarray}
\nabla &=& {{\bf u}'} \  \frac{1}{{\Phi   \left( {\bf x}  \right)}} \left( ( {\gamma  \left( u' \right)} - 1)\frac{ {{\bf u}'} . \nabla' }{{u'}^2} +  \frac{1}{{c '}^2}  {\gamma  \left( u' \right)}  \partial_{t'} \right)   + \frac{1}{{\Phi   \left( {\bf x}  \right)}} \nabla'  \label{gradientGMLT}  \\ 
\partial_{t} &=& {\gamma  \left( u' \right)} {\Phi   \left( {\bf x}  \right)} \left( \partial_{t'} +  {{\bf u}'} .
\nabla'     \right)  
\label{partialderivtGMLT}
\end{eqnarray}
These will de applied in the next sections.\\

By adequately developing the  space-time GMLT towards energy and momentum quantities,   distinct  ---first type--- \emph{momentum-energy} GMLT 's were obtained: 
\begin{eqnarray}
{{\bf p}'}  &=&  \left(\left( {{\bf p}}_\parallel  -   \frac{E}{ {c ({\bf x} )}^2} {\bf u}(\mathbf{x}) \right)   {\gamma  \left( u, {\bf x} \right)}   + {{\bf p}}_\perp \right) {\Phi   \left( {\bf x}  \right)}   \\
{E'} &=& \left( E - {{\bf p}} . {\bf u} (\mathbf{x})   \right)  \frac{{\gamma  \left( u, {\bf x} \right)}}{{\Phi   \left( {\bf x}  \right)}}
\end{eqnarray}
and its inverse;
\begin{eqnarray}
{\bf p} &=&  \left(\left( {{\bf p}'}_\parallel - \frac{E'}{ {c'}^2} {\bf u}' \right) {\gamma \left( u' \right)} + {{\bf p}'}_\perp \right) \frac{1}{\Phi \left( {\bf x}  \right)} \\
E &=&  \left( E' - {{\bf p}'} . {\bf u}' \right)  \gamma \left( u' \right)  \Phi \left( {\bf x}  \right)  \label{EnergyMomentumGMLT}
\end{eqnarray}
Now letting a spatially and kinematically coincident affected observer $S'$, with a particle with rest mass $m_0'$, attribute standard special relativistic energy, $E' = m'_0 {c'}^2$, and  zero momentum, the  corresponding expressions by the unaffected observers  $S_0$ can be obtained. The  spatial and kinematical coincidence of the $S'$  frame and the particle at $\mathbf{x}$ fixes  the frame velocity $\mathbf{u}(\mathbf{x})$ to the particle velocity $\mathbf{v}$, both attributed by $S_0$. Then  momentum, energy and mass  expressions attributed by $S_0$ are  readily obtained from the energy momentum GMLT: 
\begin{eqnarray}
{{\bf p}} \ \equiv \ {m ({\bf x} , {\bf v})} {\bf v} &, & \ \ \ E  \ \equiv \  {m ({\bf x} , {\bf v})} {c ({\bf x} )}^2  \ \ , \ \ \ \  \ {m ({\bf x} , {\bf v})} \equiv   \  m_0 ({\bf x} )   \gamma  \left( v, {\bf x} \right)  \  \equiv   \  {m_0'} \frac{{\gamma  \left( v, {\bf x} \right)}}{{\Phi   \left( {\bf x}  \right)}^3}  
\label{momentumandenergy}
\end{eqnarray}
where now  velocity $\mathbf{v}$  of  the particle at ${\bf x}$  replaces the  frame velocity $u({\bf x})$ in the $\gamma$ function;
\begin{eqnarray}
{\gamma  \left( v, {\bf x} \right)} =    \left( 1 - v^2/c(\mathbf{x} )^2 \right)^{-1/2}. 
\end{eqnarray}
eq. (\ref{momentumandenergy})c shows that in $S_0$ mass  depends on location in the gravitational field, in line with a Machian effect.\\ The momentum equation (\ref{momentumandenergy})a  can be easily reverted using mass expression (\ref{momentumandenergy})c to yield the  kinematic affecting  $\gamma$ as a function of  momentum $p$:
\begin{eqnarray}
\gamma (p, {\bf x} ) &=& \left(1 + p^2/ {m_0 ({\bf x} )}^2 c({\bf x} )^2\right)^{1/2} \label{gammaofp}
\end{eqnarray}
Using the expression $\gamma (p, {\bf x})$ in the expression  for the energy eq. (\ref{momentumandenergy})b, we can obtain the energy in terms of the canonical Hamiltonian variables; $E = E ({\bf x}, {\bf p})$.  Then  $S_0$ attributes to  particles and photons respectively the Hamiltonian energy expressions (static fields):
\begin{eqnarray}
 E   = \left( {m_0 ({\bf x} )}^2  {c ({\bf x} )}^4 + p^2 c ({\bf x} )^2\right)^{1/2}     &,&   E  = p {c ({\bf x} )}   \label{hamiltonians}
\end{eqnarray}
These Hamiltonian expressions for static source will be completed for kinematical source effects in Section (\ref{sweep}).\\
The transformation relation between two affected observers ---denoted \emph{second} type GMLT--- were extensively developed in previous work \cite{Broekaert2005a}.   Specifically, the second type GMLT's are obtained  by composing  two first type GMLT's,  $S'_1$ to $S_0$ and $S_0$ to $S'_2$, with  elimination of the  common $S_0$ perspective. 
The second type GMLT  have  an unspecified group symmetry along the frame-kinematical plane $\{{\mathbf u}_{12}, {\mathbf u}_{21}\}$, and orthogonal to it cause a simple isotropic scaling. 
When  $S'_1$ and $S'_2$ are locally coincident the second type GMLT trivially reduces to a  standard Lorentz transformation ($\Phi_1 = \Phi_2$).  The local Minkowskian metric is thus recovered in the affected perspective. 

The second type GMLT can be applied in the verification of the gravitational effects in the affected perspective, \emph{i.e.} in terms of observable variables. These relations  also describe the resulting `curved' metric of the renormalized space, which is due to the affecting of measuring rods and clocks.  \\
We note that solving  a mechanical problem does not necessitate the affected observer's perspective. Solutions can be obtained in the unaffected $S_0$ perspective and subsequently transformed by a first type GMLT into the affected perspective $S'$. In fact this corresponds to the practice in GRT, where the  dynamics is expressed in  coordinate space perspective, corresponding to  $S_0$.  While these results should be transformed again to local coordinates (corresponding to $S'$), this is usually not done  because local coordinates do only interfere at higher order relative to the post-Newtonian approximation (e.g. \cite{Rindler1979}, p 142 or \cite{MTW1973}, eq. 40.14). Adopting that same strategy to our present aim, the 1-PN equivalence of the present L-P model with GRT will be established in the coordinate perspective of $S_0$. Exceptionally the implementation of the Poincar\'e Principle of Relativity will require a transformation to the affected perspective (Section \ref{PPR}). 

\section{Sources and gravitational fields. \label{Newtonfit}}
The physical content of the model  is  fixed  by adjusting the \emph{scaling} function $\Phi$ to the limit of Newtonian gravitation. In that limit, it was shown in previous work \cite{Broekaert2005a}, that $\Phi$ obeys a static field equation.      Formally the equation  was obtained from the Newtonian energy  fit by $S_0$, with potential energy  increment according $d E = \Phi d U_{Newt.}$. This means that for $S_0$   the static energy change ---amounting to an affected increment of Newtonian potential energy---  is described identically  as the affecting of rest mass itself. The latter leading to $d E = {m_0'} {c '}^2 d \Phi$ and  subsequently the identification of $\Phi$.  This fitting procedure leads to the static field equation:
\begin{eqnarray} 
  \Delta  \Phi  \ = \ \frac{4 \pi G' }{{c '}^2} \rho({\bf x}  ) \Phi  +  \frac{\left(\nabla \Phi \right)^2}{\Phi} , &&    \Phi  \ \equiv \ \exp (\varphi),  \ \varphi \ = \  { - \frac{ G'   }{ {c '}^2}  \int_S \frac{\rho({\bf x}^*)  }{\vert {\bf x} - {\bf x}^* \vert} d^3 x^*}  
 \label{statgeneralPhi} 
\end{eqnarray}
For a  Schwarzschild configuration of a spherically symmetric source of radius $R$ and, mass $M$ the field is given by:
\begin{eqnarray}
 \Phi ({r})  \ = \ \exp \left({ - \frac{\kappa}{r} }\right), &&  r > R , \ \ \   \kappa \equiv - \frac{G' M}{{c '}^2}  \label{SSSPhi}
\end{eqnarray}
The closed form $\Phi ({r})$, developed at $O(\kappa^3/r^3)$, was shown to adequately reproduce the GRT result for {\it i)} the deflection of light by the Sun, {\it ii)} the precession of orbital perihelia, {\it iii)} the gravitational delay of radar echo, and  in $S'$ perspective; {\it iv)} the gravitational redshift of spectral  lines. 
\subsection{Kinematic gravitational source \label{sweep}}
In the static configuration the  affected fixed observer  is considered at rest to the gravitational source, and  the velocity of the test particle is considered relative to the source.
In the kinematic configuration the source has a proper movement relative to some `preferred'  frame  and ---in our model--- causes a \emph{sweep velocity}  field, denoted $\mathbf{w}$, relative to that frame.  (We  consider  `preferred'  the attribute of any given frame relative to which the gravitational source has a significant or attributable velocity.)  A test particle at location $\mathbf{x}$ has an associated \emph{effective} velocity composed of  its velocity ${\mathbf v}_w$ relative to the `preferred'  frame, now  denoted $S_w$, and, the local sweep velocity $\mathbf{w} (\mathbf{x})$.  Evidently the description of dynamics of a test particle by the affected observer $S'$, \emph{uniformly co-moving  with the unaccelerated source}, should coincide with a stationary field description. \emph{I.e.} that is what the Poincar\'e Principle of Relativity ---which Poincar\'e originally intended for ``absolute" motion---  should imply for the jointly moving source and affected observer system. Foremostly we should verify whether, for kinematic sources, the sweep velocity field correctly renders the  PN acceleration expressions of  GRT. Subsequently  these expressions should be transformed from the preferred frame $S_w$ to an affected observer  $S'$, relatively at \emph{rest} to the source.  Then, following the Poincar\'e Principle of Relativity, we need to verify whether the affected acceleration $\mathbf{a}'$ of the particle  is retrieved by the uniformly co-moving observer $S'$. This expression should then corresponds to the unaffected  acceleration  $\mathbf{a}$ for \emph{static} source, as the observer remains relatively at rest to the source.\\

First we make an educated consideration about the nature of the sweep velocity field and its effect on the velocity of a test particle.  We consider the following situation: an extended source with variable velocity distribution causes one single resultant gravitational field, and thus also one resultant affecting potential and gamma factor  on a given test particle. What is the effective velocity ---relative to the frame in which the movement of the sources are described---   that should be considered in the kinematical description of the test particle?   The velocity relative to different source parts  will in general contribute  differently. 
Each part will contribute {\emph i)} proportionally to local source density $ \rho $ (no contribution for $\rho \to 0$), \emph{ ii)} proportionally to local  source velocity  $\mathbf{v}_\rho$ (particle moves relatively at $-\mathbf{v}_\rho$), \emph{ iii)} attenuated by  distance (source contributes proportional  to  relative potential energy, at $1/r$).   Then  the resultant velocity relative to the sources is  the $\varphi$-average over the source's velocity distribution, with an as yet to be fixed parameter  $\lambda$ for amplitude fitting; 
\begin{eqnarray}
{\mathbf  w} (\mathbf{x}, t) \approx \lambda  \left< \rho (r^*) \frac{{\mathbf  v}_{\rho} (x^*)}{|{\mathbf  x - x^*}|} \right>  \label{wintuitive}
\end{eqnarray} 
We must now fix the effect of the induced velocity ${\mathbf  w}$ on a \emph{moving} particle at $(\mathbf{x},t)$.  
To order $O(w)$,  we can claim that the unaffected observers $S_w$ (with kinematic source) and $S_0$ (without kinematic source), due to their nature of being unaffected, relate their respective observed velocities by adding up classically the sweep velocity to the latter. \emph{I.e.}  $S_w$ and  $S_0$ relate the observed velocities  formally by \emph{Galilean}  composition with the sweep velocity field.
Because the sweep velocity enters as an external (potential) field, this component will not be available as a kinematical variable of the  particle's mechanics. It will  therefor  be the  velocity $\mathbf{v}_{0}$ by the unaffected observer $S_0$ which  appears as the effective velocity  in the dynamics (Subsection \ref{PN}):
\begin{eqnarray}
\mathbf{v}_0 \ &\equiv&{\mathbf v}_w \ominus {\mathbf w}  \ =_{ \vert_{\rm Galilei}}  {\bf v}  - {\bf w}   \ \ \ , \ \ \  ({\rm photon}\  \mathbf{v} = \mathbf{c}) \label{effectivecomposition}
\end{eqnarray}
 The precise nature of the dynamically  effective velocity ${\mathbf v}_{0} = {\mathbf v}_w \ominus {\mathbf w}$ to higher order will depend on the cosmological model and the gravitational field equation: closer to the source its effect dominates, and the test mass is moving increasingly relative to the rest frame of the source instead of the rest frame of the remainder matter of the universe (\cite{Dicke1957}, eqs. (5, 6)). Till $O (w)$ this competition between the frame  of the source  and ``background" should be correctly quantified by tuning the parameter $\lambda$ in expression (\ref{wintuitive}), this will be done in  Section \ref{PPR}. In the present work there will be no development beyond $O(w)$. Notice that this relation (\ref{effectivecomposition}) does not imply a change of frame ---as its Galilean form would suggest---, it describes a change of physical configuration.\\
%Also notice that the velocity of light  relative to $S_0$ will similarly depend on the sweep velocity field. In the affected perspective there remains of course  the fact that the velocity of light is the maximal velocity of physical phenomena, thus causing no tachyonic implications. The kinematical affecting factor is modified accordingly: $\gamma (v_0, \mathbf{w},  \mathbf{x} ) \equiv (1 - v_0^2/c_{0}^2)^{-1/2}$.\\
We will thus hypothesize that for moving gravitational source;\\
\indent \emph{i)} An induced velocity field, ${\mathbf  w}$, follows the field equation:
\begin{eqnarray}
\Delta {\mathbf  w} &=&  4 \pi \lambda \frac{G'}{{c'}^2} {\rho {\mathbf v}_\rho}({\mathbf x}, t)\label{wequation}
\end{eqnarray}
which includes solutions of the type eq. (\ref{wintuitive}).
This basic equation is considered approximative;  without  gravitational self-field  of the source nor propagation retardation. Also  the contribution of the field's momentum and energy density  is ignored in the present approximation. The numerical parameter $\lambda$ will be fixed by the Poincar\'e Priciple of Relativity (Section \ref{PPR}). For a rigid source, with velocity $\mathbf{u}$ relative to $S_w$, the field equation (\ref{wequation}) for $\mathbf{w}$ solves to:
\begin{eqnarray}
\mathbf{w} (\mathbf{x}) &=& \lambda \mathbf{u} \varphi(\mathbf{x})  \label{rigidsweep}
\end{eqnarray}
where $\varphi$ is given by eq. (\ref{statgeneralPhi})c. (which can be easily seen by setting the ansatz, $\mathbf{w} =  \mathbf{u} f $, in eq. (\ref{wequation}))\\
 In the given approximation, a linearly moving source is thus indeed enveloped by an instantaneous and  parallel induced velocity field which attenuates   inversely  proportional  to distance.\\
\indent  \emph{ii)}  The  space and time intervals of the unaffected observer $S_w$ (with  sweep field),  are related to the intervals of the  unaffected observer $S_0$ (without sweep field),  at $O(w^2)$, by a  ``\emph{Galilean}" space and time transformation: 
\begin{eqnarray}
\delta \mathbf{x}_0  &=&  \delta \mathbf{x}_w - \mathbf{w} \delta t_w      \\
\delta t_0  &=& \delta t_w    \label{SpaceTimeGT}
\end{eqnarray}
and
\begin{eqnarray}
\nabla_0&=& \nabla_w \\
\partial_{t_0}  &=&      \partial_{t_w}  + \mathbf{w}.\nabla_w   
 \label{gradientGT} 
\end{eqnarray}
This results in the required $\delta x_0/ \delta t _0 = v_{0} = \mathbf{v}_w - \mathbf{w}$ velocity relation eq. (\ref{effectivecomposition}). 
From the space and time relation between $S_0$ and $S_w$ an energy and momentum relation can be fixed as well. The Hamiltonian expression of a particle in the sweep velocity field can then be  considered to be the energy attributed by $S_0$, kinematically boosted by  the sweep velocity.  The composition of energy and momentum in the velocity sweep field will again be in  Galilean  approximation (\ref{effectivecomposition}):
\begin{eqnarray}
{\bf p}_0 & \equiv  & {\bf p}_w \ominus m {\bf w} \ =_{ \vert_{\rm Galilei}}  {\bf p}_w - m {\bf w} \\
 E_0 & \equiv  &  E_w \ominus {\bf p}_w. {\bf w}   \ =_{ \vert_{\rm Galilei}}   E_w  - {\bf p}_w. {\bf w}
\label{EnergyMomentumGT}
\end{eqnarray}

\subsection{Dynamics relative to moving source \label{PN}}
The time-dependent induced velocity field  must appear in the dynamics of a particle. Moreover, the particle's energy is now directly dependent on the induced velocity field. And, as is well known from standard theory of mechanics with \emph{time-variable potentials}, the energy of the particle is not conserved.  \\ 
We let the unaffected observer, $S_w$, attribute form-invariant expressions of energy and momentum according the static field expressions:
\begin{eqnarray}
E_w \ = \ m_0(\mathbf{x}) \gamma (v_w, \mathbf{x}, \mathbf{w}) c_w(\mathbf{x}, \mathbf{w})^2 &,&\mathbf{p}_w \  = \   m_0(\mathbf{x}) \gamma (v_w, \mathbf{x}, \mathbf{w})  \mathbf{v}_w  
\end{eqnarray}
with
\begin{eqnarray}
 \gamma (v_w, \mathbf{x}, \mathbf{w}) \  = \ \left(1-\frac{v_w^2}{c_w(\mathbf{x}, \mathbf{w})^2}\right)^{-1/2}.
\end{eqnarray}
While for photons:
\begin{eqnarray}
E_w \ = \ p_w c_w(\mathbf{x}, \mathbf{w}) &,& \mathbf{p}_w \  = \  \frac{E_w}{c_w(\mathbf{x}, \mathbf{w})}  \mathbf{1}_{c_w}  
\end{eqnarray}
The $S_w$ expressions do not render explicit the role of the induced velocity field in the particle's velocity $\mathbf{v}_w$. In the description the sweep field acts as a `vectorial potential', not a dynamical variable. All expressions by $S_w$ must therefor be recast on $S_0$ expressions using eqs. (\ref{SpaceTimeGT}, \ref{EnergyMomentumGT}).
E.g. the energy attributed by $S_0$ in terms of velocity ${\bf v}_0$ is  given by:
\begin{eqnarray}
E_{0} &=& m_0 (\mathbf{x}) \gamma ({\bf v}_0, \mathbf{x} ) c^2 \ = \ m_0 (\mathbf{x}) \gamma ({\bf v}_w \ominus {\bf w}, \mathbf{x} ) c (\mathbf{x})^2 \label{Eo}
\end{eqnarray}
This expression allows to discern and isolate the effect of the sweep velocity.  
Thus $S_w$ attributes ---at $O(w)$ due to the approximative relation eq. (\ref{EnergyMomentumGT})b---  an energy $E_w$  to a particle in the sweep velocity field, using the $E_0$ expression eq. (\ref{Eo}) in the energy transformation eq. (\ref{EnergyMomentumGT})b  between $S_0$  and $S_w$:
\begin{eqnarray}
E_w (\mathbf{v}_w,   \mathbf{x}) =   m_0 (\mathbf{x}) \gamma (\mathbf{v}_w - \mathbf{w}, \mathbf{x}) c(\mathbf{x})^2 +  {\bf p}_w. {\bf w} (\mathbf{x}) 
\end{eqnarray}
Now the momentum expression eq. (\ref{gammaofp}) of the $\gamma (p, \mathbf{x})$ function is used to express the energy $E_w (\mathbf{v}_w,   \mathbf{x})$ as the Hamiltonian $H(\mathbf{p}, \mathbf{x})$ of the particle  as (with $ {\bf p} \equiv {\bf p}_{w}$) till $O(w)$:
\begin{eqnarray}
H &= &m_0 (\mathbf{x}) \gamma(p, \mathbf{x}) c(\mathbf{x})^2 + {\bf p} . {\bf w}(\mathbf{x})      \label{sweepparticlehamiltonian}
\end{eqnarray}
since for $\gamma(p, \mathbf{x})$ we find that; 
\begin{eqnarray}
\nabla_p \gamma (p, \mathbf{x}) & = &  \gamma (p, \mathbf{x})^{-1}   \frac{\mathbf{p}}{m_0 ({\bf x})^2 c({\bf x})^2}
\end{eqnarray} 
and by using the  Hamilton equation for velocity (with  $\dot{\mathbf{x}} \equiv {\bf v}  \equiv {\bf v}_{w} $);
\begin{eqnarray}
{\bf v} & \equiv & \nabla_p H \  = \ \frac{{\bf p}}{m ({\bf x}, {\bf p})}   +  {\bf w}  (\mathbf{x})  \label{velocityhamiltonequation}
\end{eqnarray}
where  we used;
\begin{eqnarray}
m  ({\bf x}, {\bf p}) &=& m_0 (\mathbf{x}) \gamma(p,\mathbf{x}) \label{mofp}
\end{eqnarray}
 as in eq. (\ref{momentumandenergy})c. Notice thus  that  the momentum's canonical conjugate velocity is $\mathbf{v}$ while the \emph{effective}  dynamical velocity  is  ${\bf v} - {\bf w}$, for 
\begin{eqnarray}
{\bf p} = m  ({\bf x}, {\bf p}) ( {\bf v} - {\bf w})  \label{sweepmomentum}
\end{eqnarray}  
then;
\begin{eqnarray}
\gamma({\bf p}, {\bf x}) &=& \gamma( {\bf v} - {\bf w}, {\bf x}) \label{gammapisgammavzero}
\end{eqnarray}
 as is indeed required  to express $E( {\bf v}, {\bf x})$, eq. (\ref{Eo}), as the Hamiltonian $H( {\bf p}, {\bf x})$, eq. (\ref{sweepparticlehamiltonian}). We emphasize that, given the order of validity $O(w)$ of the Hamiltonian eq.(\ref{sweepparticlehamiltonian}),  the ensuing results of calculations should in the end always be restricted to expressions to first order in $\mathbf{w}$.  \\

\noindent For photons the description in the velocity sweep field, with  ${\bf c}_0 =  {\bf c}_w \ominus {\bf w}$ and   $\vert {\bf c}_0 \vert = c$,  leads to the Hamiltonian at order  $O(w)$:
\begin{eqnarray}
H &= &  p c (\mathbf{x}) + {\bf p} . {\bf w} (\mathbf{x})   \label{sweeplighthamiltonian}
\end{eqnarray}
and with the  Hamilton equation for velocity ($\dot{\mathbf{x}} \equiv {\bf v}_c  \equiv {\bf c}_{w} $)
\begin{eqnarray}
{\bf v}_c \  =  \ c (\mathbf{x}) {\bf 1}_p  +  {\bf w} (\mathbf{x})
\end{eqnarray}
Where we have indicated by index $c$ that ${\bf v}_c$ is the velocity of light  relative to $S_w$.

\subsubsection{Particle dynamics}
We study now the dynamics according the  Hamiltonian, eq. (\ref{sweepparticlehamiltonian}), of a test body in the gravitational fields $\Phi$ and $\mathbf{w}$ of a moving source.  From the Hamiltonian expression we immediately  obtain the classical mechanical equations of motion in coordinate space $S_w$:
\begin{eqnarray}
 H = m c^2 + {\bf p}. {\bf w}\ \ \to \ \ \dot{ {\mathbf p} } \ =\ -\nabla H, \ \ \ \ \ \ \ \dot{ {\mathbf x} } \ = \  \nabla_p H, \ \ \ \ \ \ \  \dot{ H } \ = \ \partial_t H 
\end{eqnarray}
where $\nabla \equiv \partial_{x_i} \mathbf{1}_i$ and  $\nabla_p \equiv \partial_{p_i} \mathbf{1}_{p_i}$ are the usual shorthand notations for the three dimensional partial derivatives towards $\mathbf{x}$ and $\mathbf{p}$ respectively.
For lucidity of notation we drop from now on the variable dependencies of $\gamma, {\mathbf w} , m, \varphi, \Phi$  and $c$ in our discussion of the Hamilton equations.\\
 The velocity Hamilton equation was explicited before, eq. (\ref{velocityhamiltonequation}), with $\dot{ {\mathbf x} } \equiv  {\mathbf v} $; 
\begin{eqnarray}
  {\mathbf v}  &=&   \frac{\mathbf{p}}{m} + \mathbf{w} 
\end{eqnarray}
The power Hamilton equation gives;
\begin{eqnarray}
\dot{ H } &=& \partial_t H \ = \  m c^2 \partial_t \varphi + \mathbf{p}. \partial_t \mathbf{w}  \label{powerhamiltonequation}
\end{eqnarray}
As mentioned above energy conservation is only present for time invariant potentials.   That is, no energy transfer between particle and  sweep velocity field will occur for stationary sweep velocity fields, e.g. from an axially symmetric source with constant axial rotation (cfr. concluding section \ref{conclusion}). \\
The force Hamilton equation  finally is found to be, to order $O(w)$;
\begin{eqnarray}
\dot{ {\mathbf p} } \ =\ -\nabla H &=&  - m c^2 \left(2-\frac{1}{\gamma^2}\right) \nabla \varphi - \mathbf{p} \times (\nabla \times \mathbf{w} ) - (\mathbf{p}.\nabla)\mathbf{w} \label{dyngravforce} 
\end{eqnarray}
where we have used eq. (\ref{mofp}) for $m$, eq. (\ref{gammaofp}) for $\gamma$, eq. (\ref{VSL}) for $c$ and eq. (\ref{statgeneralPhi})b for $\Phi$ in calculating; 
\begin{eqnarray}
- \nabla \left( m c^2 \right) & = & -  m_0'  {c'}^2 \nabla  \left( \Phi \left(1+ \frac{p^2 \Phi^2}{{m_0'}^2 {c'}^2 } \right)^{1/2} \right) \ = \  - m   {c}^2 \nabla \varphi - m  {c}^2    \gamma^{-2}  (\gamma^2 -1)\nabla \varphi
\end{eqnarray}
and standard calculus gives (${\bf p}$ is an independent Hamiltonian variable);
\begin{eqnarray}
 \nabla  \mathbf{p}. \mathbf{w}  & = &   \mathbf{p} \times (\nabla \times \mathbf{w} ) + (\mathbf{p}.\nabla)\mathbf{w} 
+ \mathbf{w} \times (\nabla \times \mathbf{p} ) + (\mathbf{w}.\nabla)\mathbf{p}  \ = \     \mathbf{p} \times (\nabla \times \mathbf{w} ) + (\mathbf{p}.\nabla)\mathbf{w}
\end{eqnarray}
In order now to obtain the acceleration, $\mathbf{a} \equiv \dot{\mathbf{v}}$, of the particle in the coordinate space of $S_w$, we must take the complete time derivative of the velocity Hamiltonian equation (\ref{velocityhamiltonequation}):
\begin{eqnarray}
\mathbf{a} &=& \frac{\dot{\mathbf{p}}}{m} -\frac{\dot{m}}{m^2}\mathbf{p} +\dot{\mathbf{w}} \label{protoacceleration}
\end{eqnarray}
using now again eq. (\ref{mofp}) for $m$, eq. (\ref{gammaofp}) for $\gamma$, eq. (\ref{VSL}) for $c$ and eq. (\ref{statgeneralPhi})b for $\Phi$ we obtain;
\begin{eqnarray}
\dot{m} &=&  m \left(\frac{\dot \gamma}{\gamma} - 3  \dot{\varphi} \right)\\
\dot{ \gamma} &=&  \frac{1}{\gamma} \left((\gamma^2 -1) \dot{\varphi} + \frac{ \mathbf{p} .\dot{\mathbf{p}} }{m_0^2  c^2 } \right)
\end{eqnarray}
In the latter expression for $\dot\gamma$ the  gravitational force $\dot{\bf p}$ should be substituted by its  expression eq. (\ref{dyngravforce}).\\
All required terms have been calculated now in order to explicitly render the acceleration $\mathbf{a}$, eq. (\ref{protoacceleration}).  Respecting the validity of the Hamiltonian at $O(w)$,  all  terms appearing in $\mathbf{a}$ should finally be restricted to this order. The terms  $\gamma$,  eq. (\ref{gammapisgammavzero}), and ${\bf p}$, eq. (\ref{sweepmomentum}), must be expressed again as $\gamma (v)$ and $m ({\bf v} - {\bf w})$ and are approximated to their first post-Newtonian value. The time derivatives of the fields, occuring in the expression of  ${\bf a}$, are expanded $\dot \varphi = \partial_t \varphi + \mathbf{v}.\nabla \varphi$ and $\dot {\bf w} = \partial_t  {\bf w} +  ({\bf v}.\nabla) {\bf w}$.  Then, upon gathering all parts in $\mathbf{a}$, eq. (\ref{protoacceleration}), at order $O(w, {c'}^2 \nabla \varphi^2, v^2 \nabla \varphi, v\partial_t \varphi);
$\begin{eqnarray} 
\frac{\dot{\mathbf{p}}}{m}  &\approx& -  {c'}^2 \nabla( \varphi + 2  \varphi^2)  -   v^2 \nabla \varphi - \mathbf{v} \times (\nabla \times \mathbf{w} ) - (\mathbf{v}.\nabla)\mathbf{w} \\
 -\frac{\dot{m}}{m^2}\mathbf{p} &\approx& -\left(   -  \mathbf{v}.\nabla \varphi    - 3  (\partial_t \varphi + \mathbf{v}.\nabla \varphi) \right)\mathbf{v}\\
 \dot{\mathbf{w}} &=& \partial_t  {\bf w} +  ({\bf v}.\nabla) {\bf w}
\end{eqnarray}
the expression for the acceleration $\mathbf{a}$ for  a test body in the gravitational fields of a moving source, again at order $O(w, {c'}^2 \nabla \varphi^2, v^2 \nabla \varphi, v\partial_t \varphi)$, is found to add up to:
\begin{eqnarray}
\mathbf{a} &=& -{c'}^2 \nabla(\varphi + 2 \varphi^2)  -{v}^2 \nabla \varphi+ 4 \mathbf{v} \mathbf{v}.\nabla \varphi - \mathbf{v} \times (\nabla \times \mathbf{w} ) + 3 \mathbf{v}\partial_t \varphi + \partial_t \mathbf{w} \label{particleacceleration}
\end{eqnarray}
This is exactly the \emph{form} of the 1-PN expression of the acceleration of test particle in the  gravitational fields of a moving source in General Relativity Theory (\cite{Weinberg1972}, eq. (9.5.3) till $O(\tilde{v}^6$)). Of course in GRT the modus to obtain this result is quite different;  it follows from the condition for \emph{geodesic} motion in curved spacetime.\\
Now to obtain quantitative equivalence with the acceleration  expression in GRT, the ``vector potential"  $\mbox{\boldmath$\xi$}$  from GRT (\cite{Weinberg1972}, eq. (9.1.61)) must equal minus the sweep velocity field of our present L-P model, \emph{i.e.}   $ \mbox{\boldmath$\xi$} = -\mathbf{w}$ . Then the correspondence with GRT is exact.  We will obtain, in section \ref{PPR}, this correspondence  by implementation of the Poincar\'e Principle of Relativity in our model;  from this principle the velocity sweep field $\mathbf{w}$ will be shown to satisfy the same equation as  the vector potential $\mbox{\boldmath$\xi$}$ of GRT.

\subsubsection{Photon dynamics}
The Hamiltonian (\ref{sweeplighthamiltonian}) for a photon in the gravitational fields $\Phi$ and $\mathbf{w}$ of a moving source, will lead again to the description of its motion in coordinate space $S_w$. We follow the same line of argument as for particles.\\
The classical mechanical equations  following  the photon Hamiltonian (\ref{sweeplighthamiltonian})  are, with $\mathbf{v}_c \ \equiv \ \dot{\mathbf{r}}$;
\begin{eqnarray}
H \ = \  pc + {\bf p}.{\bf w} \ \  &\to& \ \ 
\left\{\begin{tabular}{ccccl }
$\dot{\mathbf{p}} $ &=&$ - \nabla H $&=& $- p \nabla c - \mathbf{p}\times(\nabla \times \mathbf{w}) - ( \mathbf{p}.\nabla) \mathbf{w} $\\
$\mathbf{v}_c $ &=&  $\nabla_p H$ &=& $ c \mathbf{1}_p + \mathbf{w} $ \\
$\dot{H} $ &=& $\partial_t H$ &=& $2 p c \partial_t \varphi+  \mathbf{p} . \partial_t \mathbf{w}$
\end{tabular} \right.  \label{photonhamiltonequations}
\end{eqnarray}
%\begin{eqnarray}
%\dot{\mathbf{p}} &=& - \nabla H \ = \ - p \nabla c - \mathbf{p}\times(\nabla \times \mathbf{w}) - ( \mathbf{p}.\nabla) \mathbf{w} \\
%\mathbf{v}_c \ \equiv \ \dot{\mathbf{r}} &=&  \nabla_p H \ =  \  c \mathbf{1}_p + \mathbf{w} \\
%\dot{H} &=& \partial_t H \ = \  2 p c \partial_t \varphi+  \mathbf{p} . \partial_t \mathbf{w}
%\end{eqnarray}
 where   $ \mathbf{1}_p = {\bf p}/p$ is the direction of the photon momentum. We will denote the direction of propagation as $\mathbf{1}_{c}  =  \mathbf{v}_{c}/v_c $. \\
The  photon acceleration ${\bf a}$, ${\bf a} \equiv \dot{\bf v}_c$, is obtained by time derivation of the velocity Hamilton equation (\ref{photonhamiltonequations})b:
\begin{eqnarray}
\mathbf{a} &=&   {\dot{c}} \frac{\mathbf{p} }{p}  + c   \frac{\dot{\mathbf{p}}}{p} - c \frac{\dot{p}}{p}  \frac{\mathbf{p}}{p}   + \dot{\mathbf{w}} \label{protophotonacceleration}
\end{eqnarray}
In order now to express the acceleration in terms of $\mathbf{1}_{c} $, instead of $\mathbf{1}_{p}$, we must relate both. From the velocity Hamilton equation  we obtain by the scalar product with $\mathbf{1}_{c} $:
\begin{eqnarray}
  v_c  &=& c + \mathbf{w}. \mathbf{1}_{c}
\end{eqnarray}
The velocity Hamilton equation can then be reverted:
\begin{eqnarray}
 \mathbf{1}_p = \mathbf{1}_{c}  - \frac{ \mathbf{w}}{c}+ \frac{ \mathbf{w}}{c}.\mathbf{1}_{c} \mathbf{1}_{c}  \label{directionreverted}
\end{eqnarray}
%From the time derivative of $p^2 = {\bf p}^2$ we have $\frac{\dot{p}}{p} = \mathbf{1}_p. \frac{\dot{\mathbf{p}}}{p}$.\\
We  develop  all terms required in $\mathbf{a}$, eq. (\ref{protophotonacceleration}), at order $O(w, \partial_t \varphi, v_c \nabla \varphi, v_c \nabla \varphi^2)$ using again  eq. (\ref{VSL}) for $c$ and eq. (\ref{directionreverted}) for $\mathbf{1}_p$:
\begin{eqnarray} 
 {\dot{c}}  & = & 2 c'  (\partial_t \varphi + (1 + 2 \varphi)(\mathbf{v_c}.\nabla) \varphi) \\
  \frac{\dot{p}}{p}  & = &  \mathbf{1}_p. \frac{\dot{\mathbf{p}}}{p}  \ = \ - 2 c'  \mathbf{1}_{c}. \nabla \varphi -  \mathbf{1}_{c}. (  \mathbf{1}_{c}.\nabla) \mathbf{w} \\
\dot{\mathbf{w}}  & = & \partial_t   \mathbf{w}+  (\mathbf{v_c}.\nabla)  \mathbf{w} \label{dotw}
\end{eqnarray}
 Then the required terms can be written to explicitly render the acceleration $\mathbf{a}$, eq. (\ref{protophotonacceleration}), of the photon in the gravitational field, to order $O(w, \partial_t \varphi, v_c \nabla \varphi, v_c \nabla \varphi^2)$:
\begin{eqnarray}
\mathbf{a} &=& -  2 {c' }^2 \nabla (\varphi + 2 \varphi^2) + 4  {c'}^2 \mathbf{1}_{c} \mathbf{1}_{c}.\nabla (\varphi + 2 \varphi^2)  \nonumber \\
& & -   c' \mathbf{1}_{c} \times(\nabla \times \mathbf{w}) +  c' \mathbf{1}_{c}  \mathbf{1}_{c}.  (\mathbf{1}_{c}.\nabla) \mathbf{w}+  2 c' \mathbf{1}_{c}\partial_t   \varphi  + \partial_t   \mathbf{w}  \label{photonacceleration}
\end{eqnarray}
The first part of the expression is exactly the  1-PN  photon acceleration of GRT   ---till order $O(\nabla \varphi^3)$--- in the \emph{statical} configuration  (\cite{Weinberg1972}, Eq 9.2.6 and  Eq 9.2.7). What about the  correspondence with the sweep velocity dependent terms?\\
In order to compare the full correspondence with  the PN  expression of  GRT  for photon acceleration in general, with $\mathbf{w}$, we calculate ---following Weinberg's approach--- the post-newtonian development (\cite{Weinberg1972}, eq. (9.2.6))   by including in the acceleration equation (\cite{Weinberg1972}, eq. (9.1.2))  the higher order terms (\cite{Weinberg1972}, eq. (9.1.68),  eq. (9.1.69), eq. (9.1.71), eq. (9.1.73)):
\begin{eqnarray}
\frac{d {\mathbf u}}{dt} &=&  - (1+ u^2) \nabla \phi + 4  \mathbf{u} (\mathbf{u}. \nabla )\phi  \nonumber \\
&& -  (\mathbf{u} . \nabla)  \mathbf{\zeta} +   \mathbf{u} .  (\nabla) \mathbf{\zeta}  -   \mathbf{u}  \mathbf{u}.(\mathbf{u}.\nabla) \mathbf{\zeta} + 3  \mathbf{u} \partial_t \phi -  \mathbf{u}  u^2 \partial_t \phi
 - \nabla(2 \phi^2 + \psi) - \partial_t \mathbf{\zeta}  \ \ \ \ \ {\rm (GRT)} \label{GRTphotonacceleration}
\end{eqnarray}
In Weinberg's notation; $c=1$, the photon velocity is ${\mathbf u}$, and has amplitude $\vert {\mathbf u}\vert= 1 + 2 \phi + O (\tilde{v}^3)$.  \\
It can be easily shown that for $\mathbf{w} =  \lambda  \mathbf{u}  \varphi$ (our notation) and $\mathbf{\zeta}= \alpha  \mathbf{v}  \phi $ (Weinberg's  notation, $\alpha\equiv 4$), both expressions  (\ref{photonacceleration}) and (\ref{GRTphotonacceleration}) precisely correspond, e.g. for the rotation term of $\mathbf{w}$:
\begin{eqnarray}
-   \mathbf{c} \times(\nabla \times \mathbf{w})  &=& + \lambda \mathbf{u} (\mathbf{c} .\nabla) \varphi - \lambda \mathbf{u}.\mathbf{c} \nabla \varphi  \ \ \ \  {\rm (LP\ term)} \\
-  (\mathbf{u} . \nabla)  \mathbf{\zeta} +   \mathbf{u} .  (\nabla) \mathbf{\zeta}   &=& - \alpha   \mathbf{v} (\mathbf{u} . \nabla)  \phi+   \alpha \mathbf{v} . \mathbf{u}   \nabla\phi  \ \ \ \  {\rm (GRT\ term)}
\end{eqnarray}
Thus $\lambda = - \alpha$,  or $\mathbf{w} = - \mathbf{\zeta}$,  must be satisfied for precise correspondence between GRT and the GMLT approach. \\
As mentioned in  the previous subsection the requirement of the validity of the Poincar\'e Principle of Relativity, by  fixing the amplitude $\lambda$ of the sweep velocity will be shown to assure the  1-PN correspondence with GRT, for particles and photons alike. \\
As a final note on the photon dynamics, we mention that we obtain ---at order $O({\bf w})$--- using the time derivative of the photon Hamilton equation (\ref{photonhamiltonequations})a,b, expanded time derivative of ${\bf w}$ (\ref{dotw}) and, momentum and velocity relation (\ref{directionreverted}), the unit acceleration vector obeys;
\begin{eqnarray}
\dot{\mathbf{1}}_{c}
&=&  -   \left( \nabla c  \right)_{\perp{\bf 1}_c} + \frac{1}{c}  \left( \partial_t  {\bf w}  \right)_{\perp{\bf 1}_c}  -  {\bf 1}_c \times (\nabla \times {\bf w}),
\end{eqnarray}
consistent with its orthogonality requirement  ${\mathbf{1}}_{c} . \dot{\mathbf{1}}_{c} = 0$.

 \section{Affected perspective dynamics and the  Poincar\'e Principle. \label{PPR}}
Let an affected observer at rest relative to a \emph{uniformly}  moving gravitational source describe the gravitational mechanics of a test body in this field. This description should \emph{not} involve the common velocity ---of source and observer--- relative to a prefered frame.  This property in the context of relativistic movement  was set forward by  Poincar\'e as the  Principle of Relativity of movement;  it can not be observed whether one is moving, or not, \emph{uniformly} relative to a preferred frame. \cite{Poincare1902} considered it one of the main fundamental principles of physics. Note that a \emph{non}-uniform motion of a source should of course be detectable ---as in GRT for rotation and linear acceleration  \cite{Thirring1918a,Thirring1918b,LenseThirring1918,LyndenBellBicakKatz1999}. \cite{Keswani1964} has analysed the meaning of the Relativity Principle prior to 1905 and,  \cite{Torretti1983}, sec. 3.8, summarizes shortly its context. In practice Poincar\'e's approach to the Relativity Principle is mainly referenced to in philosophy and history of physics, some works by Sj\"odin and Ivert applied the principle in \cite{Sjodin1980,IvertSjodin1980}  in transformation modeling. \\
 In order to verify the absence of preferred frame effects, the mechanics of a test body should be expressed in terms of an affected observer ``locked  at a fixed distance" to the moving source. Thus after transformation from observer $S_w$ to the affected observer $S'$, at rest relative to the moving source, all kinematical reference  to a preferred frame ---as ${\bf v}$ and ${\bf w} ({\bf r})$ in particle acceleration eq. (\ref{particleacceleration})---should disappear from the mechanical description of the test body. The invisibility of a preferred  frame for  this observer  $S'$  is now settled by $\lambda$-adjustment in the field equation, eq. (\ref{wequation}),  of the sweep velocity of the source. For the  present purpose we must thus suppose the source is not accelerating relative to $S_w$; we set  $\dot{u} = 0$ in eq. (\ref{rigidsweep}). 
Since $S'$ is at rest relative to the source, the  frame velocity  of $S'$ is equal to the source velocity  relative to $S_w$, both are ${\bf u}$. The acceleration ${\bf a}'$  of a particle relative to $S'$, should not depend on $\mathbf{w}$   anymore eq. (\ref{particleacceleration}), for to $S'$ the seemingly static source should not be attributed a sweep velocity field.\\
 In the particle mechanics description  by $S_w$, the mass $m$, the momentum $p$ and  $\gamma$-factor  are dependent on $\mathbf{v}_0$, eqs. (\ref{sweepparticlehamiltonian},\ref{gammapisgammavzero}, \ref{sweepmomentum}).  In order to have this dynamical dependency in $\gamma$, it is required to use the  GMLT with effective frame velocity $\mathbf{u} - \mathbf{w}$, i.e.  $\mathbf{u}_0$. It is thus the   $S_0$ to $S'$ GMLT which assures the  kinematic affecting  and for   $\delta \mathbf{x}' = 0  \to \delta\mathbf{x}/\delta t  = \mathbf{u}_0$, we have $\delta \mathbf{x}  = \delta \mathbf{x}_0$ and $ \delta t = \delta t_0$. The space time GMLT between $S'$ and $S_0$ is:  
\begin{eqnarray}
\delta{\bf x}'  &=&  \left(( {{\delta{\bf x}_0}_\parallel} -{\bf u}_{0}  \delta t_0)  {\gamma_0}  ({\bf u}_{0} )    +
  {\delta{{\bf x}_0}_\perp} \right) \frac{1}{{\Phi   \left( {\bf x}  \right)}}  \label{SpaceSotoSac} \\ 
\delta t'  &=& \left( \delta t_0 - \frac{ {\bf u}_0 . \delta{\bf x}_0 }{{c_0}^2}  \right)  {\gamma_0} ({\bf u}_{0} )  {\Phi   \left( {\bf x}  \right)}  \label{TimeSotoSac}
\end{eqnarray} 
with $\gamma_0 = \left(1 -u_0^2/c^2 \right)^{-1/2}$ ---for $S_0$; $c_0 \equiv c$--- precisely as $\gamma$ in eq. (\ref{gammapisgammavzero}).\\ 
The relation of $S'$ with  $S_w$ must then be obtained  by composition with the Galilean relation (\ref{SpaceTimeGT}): $S_w \leftarrow^{_{\, GT}} \rightarrow S_0 \leftarrow^{_{\, GMLT}} \rightarrow S'$.  \\
The velocity relation obtained from the $S_0$ to $S'$ space and time GMLT ---relation (\ref{velocityrelation}) but with effective velocities--- is now;
\begin{eqnarray}
{\mathbf v}' & = & \frac{ {\bf v}_0 -  {\bf u}_0 + ( \gamma^{-1} -1 )   {{\bf v}_0}_\perp}{1 - {\bf u}_0.{\bf v}_0 \mathbf{c}_0 ^{-2}}  \frac{1}{\Phi ({\bf x})^2}  \label{velocityrelation0}
\end{eqnarray}
where the orthogonality is relative to ${\bf u}_0 $. The link with the velocity ${\bf v}$ is provided by Galilean $O({\bf w})$ relation between   $S_w$ and $S_0$;
\begin{eqnarray}
{\bf v}_0 &=& {\bf v} - {\bf w} \label{Galileanvelocitycomposition}
\end{eqnarray}
These velocity relations will be used subsequently to establish relations between observed accelerations in $S'$ and $S_w$.
(We recall that the velocity and acceleration $\mathbf{v}$ and $\mathbf{a}$, as in (\ref{particleacceleration}, \ref{photonacceleration}) are relative to $S_w$, the index $w$ on these quantities has been dropped for clarity.)

\subsection{Relativistic acceleration transformations. \label{ACCELERATION}}
The acceleration of particles or photons  in curved space-time is seldom used in GRT because it  depends on the chosen metric of the coordinate system.  However a \emph{transformation} of  acceleration  is well known for  Fermi and  Schwarzschild coordinates.  A number of specific acceleration transformation laws in GRT were  described by  Mishra  and Rindler  \cite{RindlerMishra1993, Mishra1994} and more generally  using generic coordinates by \cite{BiniCariniJantzen1995} and,   \cite{BunchaftCarneiro1998} for static observers (a coordinate-free space-time decomposition of a covariant expression):
\begin{eqnarray}
\mathbf{a}  & = &  \mathbf{g} -   \mathbf{v}  ( \mathbf{v}.\mathbf{g}) 
\end{eqnarray}
where $\mathbf{a}$ is the general  local 3-proper-acceleration of a  particle  in a static gravitational field  and $\mathbf{g}$  is this same acceleration but with the ``physical" relative  velocity  $\mathbf{v} = 0$.  Following the choice of adapted coordinates in which to transform the acceleration (e.g. Fermi coordinates) this equation amounts to a kinematic decomposition of the observed gravitational acceleration.\\
The specific case of the description of  ``physical"  acceleration relative to a static  observer in the Schwarzschild metric has been developed in the literature as well. For the  (local) affected radial acceleration observed in a static frame $S'$, McGruder gives in GRT \cite{McGruder1982}:
\begin{eqnarray}
a_l  &=& g \left( v_R^2 - v_l^2 -1 \right) + O (r^{-3}) \label{RadAccSchwarz}
\end{eqnarray}
with $v_R$ the affected radial velocity and  $v_l$ the affected transversal velocity (here $g = {c'}^2 \kappa/r^2$). \\
In our model, the particular appearance of the  scaling function $\Phi$ in the space-time GMLT (\ref{SpaceSotoSac}, \ref{TimeSotoSac}) requires comparison to GRT-descriptions in \emph{isotropic} Schwarzschild coordinates.   We will therefor only need to find correspondence with the affected radial acceleration, eq. (\ref{RadAccSchwarz}).\\
We recall that in GRT the acceleration of a test particle in a local inertial free frame is obtained by a coordinate transformation of the \emph{covariant} derivative of the velocity; this renders the metric connections null and the local relative acceleration  vanishes \cite{Broekaert2006}. \\ 
At present  we verify the local acceleration expressions \emph{i}) for the static case, as in eq. (\ref{RadAccSchwarz}) and, \emph{ii}) for the case of the observer tracking the uniformly moving source as requied for checking the Poincar\'e Relativity principle.\\
In these cases ---with $\dot{\bf u} = 0$--- the acceleration transformation is obtained by applying the standard time derivation to the velocity transformation (\ref{velocityrelation0})  \cite{Broekaert2005a}.  \\
The  acceleration transformation with $\dot{\mathbf{u}} = 0$ and  with  \emph{frame} acceleration attributed by $S_0$; $\dot{\mathbf{u}}_0 =  -  \dot{\mathbf{w}} $, is then given by:
\begin{eqnarray}
{\mathbf{a}'} &=&  \left(1-\frac{ \mathbf{u}_0. {\mathbf{v}_0}}{c_0^2}\right)^{-2} \frac{1}{{\gamma_0}{\Phi}^{3}} \left\{    
{\mathbf{a}_0}_\parallel    +\frac{1}{{\gamma_0}}    {\mathbf{a}_0}_\perp  +  {\mathbf{v}'}  \frac{{\Phi}^2 }{{c_0}^2 }  \mathbf{u}_0  .  {\mathbf{a}}_0  \right. \nonumber \\
& & - 2{\mathbf{v}'} {\Phi}^2 \left( 1 + \frac{ \mathbf{u}_0.{\mathbf{v}_0}}{{c}_0^2 }   \right)  
\dot{\varphi}_0       + 2  {\mathbf{v}_0}_\perp  {\gamma_0}    \frac{u_0^2}{{c}_0^2 } \dot{\varphi}_0 
 \nonumber \\
&&    -  \dot{\mathbf{w}}+ \left(\frac{1}{{\gamma_0}} -1 \right)\left(   \frac{ \left( \dot{\mathbf{w}}  \times {\mathbf{v}_0}   \right)\times  \mathbf{u}_0  }{u_0^2}  +\frac{  \left(  \mathbf{u}_0\times {\mathbf{v}_0} \right) \times  \dot{\mathbf{w}}}{u_0^2}  -2 \frac{ \dot{\mathbf{w}}. \mathbf{u}_0}{u_0^2} {\mathbf{v}_0}_\perp\right)  \nonumber \\ & & \left.   -  {\mathbf{v}_0}_\perp  {\gamma_0}  \frac{ \dot{\mathbf{w}}. \mathbf{u}_0}{{c}_0^2}     +{\mathbf{v}' } 
\frac{{\Phi}^2 }{{c_0}^2 }  \dot{\mathbf{w}} .{\mathbf{v}_0}  \right\}   \label{avectoavecac}
\end{eqnarray}
Which is complemented with  the Galilean relation between $S_0$ and $S_w$:
\begin{eqnarray}
\mathbf{a}_0  &=& \mathbf{a} - \dot{\mathbf{w}}    
\end{eqnarray}
Taking into account the validity of order $O(w)$ the composition of both transformation, with the use of the velocity decomposition (\ref{Galileanvelocitycomposition}), gives for $\mathbf{a}'$:
\begin{eqnarray}
{\mathbf{a}'} &=&  \frac{1}{{\Phi}^{3}} \left\{    
 \left(1 +2 \frac{ \mathbf{u}. {\mathbf{v}}}{c^2}\right){\mathbf{a}}   +  {\mathbf{v}'}  \frac{{\Phi}^2 }{{c}^2 }  \mathbf{u} . \mathbf{a}  - 2{\mathbf{v}'} {\Phi}^2 
\dot{\varphi}_0       \right\}  \label{atransformationsweep}
\end{eqnarray}
notice that no supplementary introduction of $\mathbf{w}$-terms occurs at this order. In order to express the component with ${\dot\varphi}_0 \equiv \partial_{t_0} + \mathbf{v}_0.\nabla_0 \varphi$ in terms of $S'$, the gradient transformation eq. (\ref{gradientGMLT}) must be used and, the  relativistic first-order approximated velocity relation $S_0$ to $S'$; 
\begin{eqnarray}
\mathbf{v}_0 &\approx&  {\mathbf v}' \left( 1+  \frac{\mathbf{u'} .{\mathbf v}' }{{c'}^2}  \right) \Phi^2  - \mathbf{u'} \Phi^2 
\end{eqnarray}
leading to the expression ${\dot\varphi}_0$ with first relativistic correction;
\begin{eqnarray}
\dot \varphi _0&\approx& \Phi \left( 1 + \frac{{\mathbf u}'.{\mathbf v}'}{{c'}^2}\right) \left( \partial_{t'} \varphi + {\mathbf v}'.\nabla' \varphi \right)  \label{dotvarphi}
\end{eqnarray} 
In order to obtain the full expression of  $\mathbf{a}'$, eq. (\ref{atransformationsweep}), in $S'$, the velocity relations  approximated  to relativistic first-order between $S'$ and $S_w$ must be used:
\begin{eqnarray}
{\mathbf v}' & \approx &  \frac{1}{\Phi^2}  \left({\mathbf v} \left(1 + \frac{{\mathbf u}.{\mathbf v}}{c^2}\right) - {\mathbf u}\right)  + O(v^2/c^2 u)     \label{vtovacwithsweep}
\end{eqnarray}
while for the photon case  an additional appearance of $\mathbf{w}$ needs to be retained:
\begin{eqnarray}
{\mathbf c}' & \approx &  \frac{1}{\Phi^2}  \left({\mathbf c} \left(1 + \frac{({\mathbf u}- {\mathbf w}).{\mathbf c}}{c^2}\right) - {\mathbf u}\right)  + O( u)     \label{vtovacwithsweepfoton}
\end{eqnarray}
Finally we can insert the acceleration expressions which were obtained form the Hamiltonian mechanics in the perspective of $S_w$. For easy reference, we apply the transformation first in the case of a static source.

\subsection{Static source ---Endorsement of the acceleration transformation}
The particle acceleration (\ref{particleacceleration}) in $S_w$, now with $\mathbf{w}=0$, can  be transformed to  the fixed observer $S'$. The acceleration transformation, eq. (\ref{atransformationsweep}), ${\bf u}' = 0, {\bf w} = 0$, and gradient-GMLT eq. (\ref{gradientGMLT}) gives in the affected observer's perspective to first Post-Newtonian order:
\begin{eqnarray}
{\bf a}'  &=&   - {c '}^2   \nabla' \varphi  -    {v'}^2  \nabla'  \varphi  +  2  {{\bf v}'} {{\bf v}'}.\nabla'  \varphi     \label{staticacceleration}
\end{eqnarray}
Similarly, the photon acceleration  in affected and unaffected perspective are related by the acceleration transformation, eq. (\ref{atransformationsweep}), with  ${\bf u}'=0$, ${\bf v}' = {\bf c}'$, $\bf{w}= 0$.  In the affected perspective of $S'$ the photon acceleration (\ref{photonacceleration}) becomes to first Post-Newtonian order:
\begin{eqnarray}
\mathbf{a}' &=&   2  \mathbf{c}' \mathbf{c}'.\nabla' \varphi  - 2 {c'}^2   \nabla' \varphi  \ = \   2  \mathbf{c}'    \times ( \mathbf{c}' \times \nabla' \varphi )\label{staticphotonaccelerationaac} 
\end{eqnarray}
and
\begin{eqnarray}
\dot{\mathbf{1}_{c'}}
&=&  2 c'\left( \mathbf{1}_{c'}   \mathbf{1}_{c'} .\nabla' \varphi -    \nabla' \varphi \right) \ = \  \frac{\mathbf{a}' }{c'}
\end{eqnarray}
endorsing the view that in the affected perspective of $S'$ light curves but does not change velocity amplitude in a gravitational field.\\
In order to compare  the L-P acceleration with  the GRT term, eq. (\ref{RadAccSchwarz}),  we   specify radial and lateral velocity $ \mathbf{v}  =  \mathbf{v}_r +  \mathbf{v}_l$, and set ${\mathbf{g}'}  \equiv -  {c'}^2 \nabla' \varphi$,  in the  acceleration ${\bf a}'$, eq. (\ref{staticacceleration});
\begin{eqnarray}
%{\mathbf{a}'}_{v'} &=&  {\mathbf{g}'}  + \frac{1}{{\Phi}^3} \left\{  -   v^2   \nabla \varphi +  2{\bf v} {\bf v}.\nabla \varphi  \right\} \ = \ {\mathbf{g}'}     -   {v'}^2   \nabla' \varphi +  2{\bf v'} {\bf v'}.\nabla' \varphi  \\
% {\mathbf{a}'}_{v'}&=&  (1 +  {v'}^2/c^2 )   {\mathbf{g}'}  - 2{\bf v'} {\bf v'}. {\mathbf{g}'}/c^2\\
{\mathbf{a}'} &=&  (1 +  {v'_r}^2/{c'}^2+  {v'_l}^2/{c'}^2 )   {\mathbf{g}'}  - 2({\bf v'}_r + {\bf v'}_l) ({\bf v'}_r + {\bf v'}_l). {\mathbf{g}'} /{c'}^2
\end{eqnarray}
With ${\bf v'}_l. {\mathbf{g}'}/{c'}^2 =0$ we have the lateral and radial components of the affected acceleration
\begin{eqnarray}
{{a}' }_r 
&=&  (1 -  {v'_r}^2/{c'}^2+  {v'_l}^2/{c'}^2 )   {g'} \\
{\mathbf{a}'}_l
&=&   - 2 \mathbf{v'}_l { v'}_r   {g}' /{c'}^2
\end{eqnarray}
Which endorses the radial acceleration expression, eq. (\ref{RadAccSchwarz}), of GRT in Schwarzschild metric.

\subsection{Kinematic source ---Implementation of the Poincar\'e Principle of Relativity }
 In the configuration required for checking Poincar\'e's Principle of Relativity,  the source moves by with a velocity $\mathbf{u}$  relative to $S_w$.  Thus to $S_w$ the field $\Phi$ varies in time; following the time-independent field equation, eq. (\ref{statgeneralPhi}), $\Phi$ changes by uniform translation with the source at velocity $\mathbf{u}$, $\varphi = \varphi (\mathbf{r}_i - \mathbf{u}_w ( t_w- {t_w}_i)  )$. The observer $S'$ is moving with frame velocity  $\mathbf{u}$ along with the source at a distance. Therefor the observer  $S'_w$ remains in a  fixed position relative to the source, and thus to its field $\varphi$:
\begin{eqnarray}
%\partial_{t'} \varphi &=& {\gamma  \left( u_0 \right)} \frac{1}{\Phi }  \left( \partial_{t_0} +  {{\bf u}_0} . \nabla_0     \right)   \varphi \\
%&=& {\gamma  \left( u_0 \right)} \frac{1}{\Phi }  \left( \partial_{t_w}  + \mathbf{w}.\nabla_w +  {{\bf u}_0} . \nabla_w     \right)   \varphi \\
\partial_{t'} \varphi &=& {\gamma  \left( u_0 \right)} \frac{1}{\Phi }  \left( \partial_{t_w}  + \mathbf{u}_w.\nabla_w   \right)   \varphi (\mathbf{r}_i - \mathbf{u}_w ( t_w- {t_w}_i)  ) = 0 \label{varphiacconstant}
%&=& 0 \label{varphiacconstant}
\end{eqnarray}
where we used the gradient transformations, eqs. (\ref{gradientGMLT}, \ref{gradientGT}), and  velocity relations of  $S'$, $S_0$ and $S_w$ to first Post-Newtonian order. \\
--- {\bf Particle dynamics}\\
%The acceleration ${\mathbf a}$  (\ref{particleacceleration}) expressed in terms of $S'$, again using %(\ref{vtovacwithsweep}, \ref{dotvarphi}),  is given by:
%\begin{eqnarray}
%{\mathbf a}   &=& -  {c'}^2 \nabla' (\varphi + 2 \varphi^2) - {v'}^2  \nabla' \varphi   + 4 \mathbf{v}' \mathbf{v}'.\nabla' %\varphi + 3 \mathbf{v}' \partial_{t'} \varphi  
%+(2+\lambda)   {\mathbf u}'.{\mathbf v}'  \nabla' \varphi   \nonumber \\
%&& -   {\mathbf v}' {\mathbf u}'.\nabla' \varphi - (4+\lambda){\mathbf u}' {\mathbf v}'.\nabla' \varphi - %(4+\lambda){\mathbf u}' \partial_{t'} \varphi  + O(u^2, u v^2/c^2)
%\end{eqnarray}
The particle acceleration ${\mathbf a}'$, in the perspective  of $S'$ fixed relative to the moving source,  is  obtained using acceleration transformation, Eqs, (\ref{atransformationsweep}), with  ${\mathbf a}$, eq. (\ref{particleacceleration}),  reformulated in terms of $S'$, using eqs.  (\ref{vtovacwithsweep}, \ref{dotvarphi}, \ref{rigidsweep}). 
The acceleration ${\bf a}'$ to first Post-Newtonian order is given by: 
\begin{eqnarray}
{\mathbf a}'   &=& -  {c'}^2 \nabla' (\varphi + 2 \varphi^2) - {v'}^2  \nabla' \varphi   + 2 \mathbf{v}' \mathbf{v}'.\nabla' \varphi
  \nonumber \\
&& + (4+\lambda) \left(  {\mathbf u}'.{\mathbf v}'  \nabla' \varphi  -  {\mathbf u}' {\mathbf v}'.\nabla' \varphi  \right) + O(u^2, u v^2/c^2)
\end{eqnarray}
where the time-independence of $\varphi$, eq. (\ref{varphiacconstant}), relative to the particular observer $S'$, was used.\\
All reference to movement  relative to the `preferred' frame should disappear: the terms containing the frame velocity $\mathbf{u}'$ should vanish. The Poincar\'e Principle thus requires;
 \begin{eqnarray} 
\lambda &=&  - 4 \label{lambdaisminusfour}
\end{eqnarray} 
 The affected observer $S'$  then attributes to the particle the acceleration  as if  in the stationary case eq. (\ref{staticacceleration}). 
With the Poincar\'e Principle implemented,  the sweep velocity is ``normalized" at $\mathbf{w} =  - 4 \mathbf{u} \varphi = 4 \mathbf{u} \kappa/r$. 
 The 1-PN  acceleration expressions of the L-P model, eqs.  (\ref{particleacceleration}, \ref{photonacceleration}),  now correspond exactly  with the acceleration expressions of GRT, because $\mathbf{w}$ formally and \emph{quantitatively} satisfies the same field equation (\ref{wequation}) of as the one for the  ``vector potential" $\zeta$ of GRT (\cite{Weinberg1972}, Eq 9.1.161).\\
---  {\bf Photon dynamics}\\
  In the affected perspective of $S'$, uniformly co-moving with the source, we should retrieve  the static  description, eq. (\ref{staticphotonaccelerationaac}), of the photon acceleration.
%The photon acceleration $\mathbf{a}$ given by (\ref{photonacceleration}), and  expressed in the perspective of $S'$ %to required order $O({u'}^2 \nabla \varphi)$ is given by:
%\begin{eqnarray}
%\mathbf{a} &\approx&  -2 {c'}^2 \nabla' \varphi + 4  \mathbf{c}'  \mathbf{c}' .\nabla' \varphi   + 2  \mathbf{c}'  \partial_{t'} %\varphi - (\lambda + 4) \mathbf{u}'  \partial_{t'} \varphi - (\lambda +4 ) \mathbf{u}'  \mathbf{c}' .\nabla' \varphi  - 2  %\mathbf{c}'   \mathbf{u}' .\nabla' \varphi  \\
%& & +  \lambda   \mathbf{u}'  \mathbf{c}'.\nabla' \varphi  +  (8-\lambda) \mathbf{c}' \frac{\mathbf{u}'.\mathbf{c}'}{{c'}^2} %\mathbf{c}' .\nabla' \varphi +  6 \mathbf{c}' \frac{\mathbf{u}'.\mathbf{c}'}{{c'}^2} \partial_{t'} \varphi
%\end{eqnarray}
 The acceleration transformation (\ref{atransformationsweep})   will render the photon acceleration ${\mathbf a}'$ in the perspective  of $S'$. With the time-independence of $\varphi$ relative to $S'$, eq. (\ref{varphiacconstant}), and   the  $S_w$ acceleration  ${\mathbf a}$ (\ref{photonacceleration}) for photons in terms of $S'$, using (\ref{vtovacwithsweepfoton}, \ref{dotvarphi}, \ref{rigidsweep}), ${\bf a'}$ is given by: 
\begin{eqnarray}
\mathbf{a}' 
&=&   -2 {c'}^2 \nabla \varphi + 2  \mathbf{c}'  \mathbf{c}' .\nabla' \varphi  + (\lambda + 4) \left(  \mathbf{u}'  \mathbf{c}'.\nabla' \varphi -  \mathbf{c}'   \mathbf{u}' .\nabla' \varphi  \right)   + O(u^2)
\end{eqnarray}
Thus, consistent with the previous particle  case,  again the normalization $\lambda = -4 $  leads to the elimination of the $u'$-terms with reference to the prefered frame.\\
 We conclude that the implementation of the Poincar\'e Principle of Relativity, by eq. (\ref{lambdaisminusfour}), establishes, till order $O(u^2)$, and both for particles and photons,  the equivalence of the GMLT-based description and GRT.

\section{Interpretation and conclusions\label{conclusion}} 
In terms of interpretation the presented  gravitation model has the advantage of retaining a classical ontology of  Hamiltonian mechanics with a gravitational field  in a flat-metric space. While the construction of the model requires a number of consistent hypotheses ---GMLT, VSL, form-invariance, PPR---  their nature is proper to a Lorentz-Poincar\'e approach to gravitation. \\

Two supplementary examples of geometric features interpreted physically in terms of the present scalar--vector model show the interpretational merits of our model.\\
\emph{Ex. 1)} The derivation of PN expressions in GRT requires the posing of the harmonic coordinate conditions in order to constrain the degrees of liberty of the GRT field equations. The time-component of the  harmonic coordinate condition (in weak field approximation, $O(\nabla\varphi^2)$) (\cite{Weinberg1972}, eq. 9.1.66) in $S_0$ perspective becomes, in view of the relation between $\mathbf{\zeta}$ and $\mathbf{w}$:
\begin{eqnarray}
4 \frac{\partial \varphi}{\partial t} + \nabla .\mathbf{\zeta} = 0 &\to&  4 \dot \varphi \vert_{v=u} = 0  \ \to \ \partial_{t'} \varphi = 0 \label{harmoniccoordinatecondition}
\end{eqnarray}
The harmonic coordinate condition expresses precisely that the observer $S'$, with frame-velocity $\mathbf{u}$, is at rest relative to the moving source. Or, taking into account the transformation for gradient operators (\ref{gradientGMLT}), the time-component of the harmonic coordinate condition is satisfied in the affected perspective when $\partial_{t'} \varphi = 0$.\\
\emph{ Ex. 2)} The Lense-Thirring effect \cite{LenseThirring1918}, or  `frame dragging',  of an orbiting system due a rotating source is interpreted in GRT as a local  deformation of the space-time continuum.  In the present model the equivalent effect is easily obtained. Let $\rho = \rho ({r'})$ and ${\mathbf \Omega}' $ fixed , $r> R$,  and   ${\mathbf J}_s$ be the angular momentum of the source, then the resultant sweep velocity field according (\ref{wequation}) (cfr \cite{Weinberg1972}, eq. 9.5.18):
\begin{eqnarray}
{\mathbf w} &=& -4     \frac{G'}{2{c'}^2} \frac{{\mathbf r} }{r^3}\times  {\mathbf J}_s, \ \ \ {\mathbf J}_s \ \equiv \ \frac{8 \pi}{3} \int_R {\mathbf \Omega}' \rho({r'}) {r'}^4 d {r'}
\end{eqnarray}
With orbital momentum ${\mathbf L} \equiv {\mathbf r}\times{\mathbf p} $, the Hamiltonian energy expression (\ref{sweepparticlehamiltonian}),  can be re-expressed with coupling of source angular momentum ${\mathbf J}_s$  and orbital momentum of testbody ${\mathbf L}$ till $O(J_s^2)$:
\begin{eqnarray}
H &=&  \left( {m_0}^2 c^4+ p^2 c^2\right)^{1/2} + 4 \frac{  G'}{2 {c'}^2 }  \frac{{\mathbf J}_s.{\mathbf L} }{r^3}  \end{eqnarray}
Straightforward development of the Hamilton-Lagrange principle  then leads to the first  acceleration correction ${\mathbf a}_1 $ of GRT at $O( v \nabla \varphi J_s/r^2, v v^2/c^2 J/r^3  )$:
\begin{eqnarray}
{\mathbf a}_1 &=&  4  \frac{ G'}{2{c'}^2 r^3}  \left(  {\mathbf v} \times {\mathbf J}_s   - 3  {\mathbf v}  \times  {{\mathbf J}_s}_r  \right) , \ \ \  {{\mathbf J}_s}_r  \equiv \mathbf{J}_s. \mathbf{1}_r  \mathbf{1}_r \nonumber
\end{eqnarray} 
These terms precisely cause the `frame dragging' of the particle frame  in geometrodynamic interpretation, while in the present scalar--vector model the effect is merely  caused by the different relative  velocities with respect to  the parts of the rotating source. Each part of the source therefore exerts a different kinematical $\gamma$-contraction and time dilation on the particle. \\
These examples and the description of particle and photon dynamics show that the interpretation of the gravitational effects  extends conservatively the Lorentz-Poincar\'e interpretation of  SRT.  \\

We recall that  we used  Poincar\'e's principle of Geometric Conventionalism as a fundamental premiss to develop this model. This principle demands that the gravitational phenomenology of GRT should be reproduced regardless the choice of the fundamental metric. The main result of this present work is that we realized this till the first Post-Newtonian order of GRT.  Whether the Poincar\'e Principle of Relativity  can be successfully  implemented beyond $O(u^2)$ needs to be studied in later work:  is the mutual elimination  of terms   at higher order   rigorous and absolute, as was hypothetically advanced by Poincar\'e? (\cite{Poincare1902}, p. 183) In the present work the model ignored gravitation propagation, source acceleration, etc. Some of these matters will be covered in upcoming work.

\section*{Acknowledgement.}
 The author wishes to thank Ing. K. Tilmant  for valuable discussions on the problem of gyroscopic precession in the Lorentz-Poincar\'e model.  Prof. Harvey Brown is  acknowledged for discussions and hospitality  at the Philosophy of Physics Group in Oxford. \\

This work was supported by FWO--Vlaanderen, project F6/15-VC. A87.

\section*{Appendix: The constrained differential nature of the space and time GMLT \label{appendix}}
\setcounter{equation}{0}
Since GRT  is a differential geometric theory of gravitation, the space time GMLT transformations, eqs. (\ref{SpaceTimeGMLTsotosac},\ref{SpaceTimeGMLTsactoso}) b, beg the comparison with a differential diffeomorphic relation connecting $S_0$ and $S'$.  The Lorentz-Poicar\'e type approach to gravitation however clearly specifies the Gravitationally Modified Lorentz Transformations to  intervals $\delta x'$  which are gravitationally \emph{scaled} infinitesimal intervals \emph{constrained by conditions for observation} by $S'$.\\
The  complete differential of the diffeomorphism $x = x (x')$ relating the coordinate systems of $S_0$ and $S'$ gives:
\begin{eqnarray}
 d x_i \ = \ \frac{\partial x_i }{\partial x_j'} d x_j' +  \frac{\partial x }{\partial t'} d t' \ &,& \ d t  \ = \    \frac{\partial t}{\partial x_j'} d x_j' +  \frac{\partial t }{\partial t'} d t'     \label{completediff}
\end{eqnarray}
which for consistency requires commutativity of the second derivatives ($\mu= 0, ..., 3$);
\begin{eqnarray}    
 \frac{\partial^2 x_\mu }{\partial x_\sigma' \partial x_\tau'}  &=&  \frac{\partial^2 x_\mu }{\partial x_\tau' \partial x_\sigma' }. \label{commutation}
\end{eqnarray}
The space and time GMLT, eqs. (\ref{SpaceTimeGMLTsactoso}) b, can be written as:
\begin{eqnarray}
\left(\begin{array}{c} \delta t \\ \delta{\bf x}  \end{array} \right)
\ = \ \Lambda^{-1}(\mathbf{u}, \mathbf{r})
\left(\begin{array}{c} \delta t' \\ \delta{\bf x}' \end{array} \right) ,    && \Lambda^{-1}(\mathbf{u}, \mathbf{r}) \equiv \left(\begin{array}{ll}
 \gamma \Phi^{-1}  &  -\mathbf{u}' {c'}^{-2} \gamma \Phi^{-1} \\
 -\mathbf{u}'  \gamma \Phi   &  \mathbf{1}  \Phi  + \frac{\mathbf{u}'_i\mathbf{u}'_j}{{u'}^2} (\gamma-1)  \Phi  \end{array} \right)
\end{eqnarray}
or in factorized form;
\begin{eqnarray}
\left(\begin{array}{c} \delta t \\ \delta{\bf x}  \end{array} \right)
& = &   \left(\begin{array}{ll}
  \Phi^{-1}  & 0 \\
 0  &    \Phi  \end{array} \right) \Lambda^{-1}(-\mathbf{u}')
\left(\begin{array}{c} \delta t' \\ d{\bf x}' \end{array} \right)  \ = \     \left(\begin{array}{ll}
  \Phi^{-1}  & 0 \\
 0  &    \Phi  \end{array} \right) 
\left(\begin{array}{c} \delta \tau' \\ \delta \mbox{\boldmath$\xi$}' \end{array} \right) 
\end{eqnarray}
where the coordinates   $(\delta \tau', \delta \mbox{\boldmath$\xi$}') $ were defined as $\Lambda(u') (\delta t', \delta \mathbf{x}')$ and $\Lambda (\mathbf{u}')$ is a standard Lorentz Transformation, with $\Lambda^{-1}(-\mathbf{u}') = \Lambda(\mathbf{u}')$.\\
 We can now concisely write the GMLT as;
\begin{eqnarray}
\delta t =  \Phi^{-1} \delta \tau' &,&  \delta \mathbf{x} = \Phi \delta \mbox{\boldmath$\xi$}' \label{simpleGMLT}
\end{eqnarray}
which simplifies the discussion; \emph{i.e.} as if the affected observer was in relative rest to $S_0$.\\
The gravitational field is claimed to affect  an infinitesimal ruler as a local isotropical scaling,  and the periods of clocks are also locally dilated. The local spatial measurements of $S'$ and $S_0$ are therefore related by a local scaling transformation.  We let the observer $S'$ and $S_0$ relate to Cartesian coordinates. An observation  of  an infinitesimal spatial  interval $\delta x'$ in direction $1_{x'}$, spanned by a material ruler, will extend over a different length along  $1_{x}$ of $S_0$ and because of the \emph{scaling} effect no other orthogonal interval in $S_0$ is covered by this ruler: $\delta y = 0$ and $\delta z = 0$. Moreover the measurement is done by simultaneously marking the ends of the infinitesimal ruler, leading to $\delta t = 0$. For the orthogonal rulers,  $\delta y'$ and $\delta z'$, similar constraints are imposed.  \\
For the observation of an infinitesimal time interval $\delta t'$  the clock is supposed to be stationary, \emph{i.e.} with no displacement occurring:  $\delta x = \delta y = \delta z = 0$. \\
 We compare these intervals to similar observations in GRT in local coordinates of the observer using;
\begin{eqnarray}
ds^2 &=& g_{\mu\nu} dx^\mu dx^\nu 
\end{eqnarray}
with similar constraints we have (we use primed symbols for the local coordinates);
\begin{eqnarray}
d\tau' {\vert_{_{dx_i =0}}} \ = \  (g_{00})^{1/2} dx^0 \ , \ d {\xi'_i}{\vert_{_{dx_{j \neq i} =0, d\tau = 0}}} \ = \  (g_{ii})^{1/2} dx^i
\end{eqnarray}
From which we obtain the relation between the GMLT intervals and the constrained differentials of GRT in the limit of $\delta \tau', \delta \xi'_i \to 0$;
\begin{eqnarray}
\delta \tau' \ = \ d\tau'{\vert_{_{dx_i =0}}}   &,&\ \delta \xi'_i \ = \  d {\xi'_i}{\vert_{_{dx_{j \neq i} =0, d\tau = 0}}} \end{eqnarray}
The GMLT does therefor not correspond to the differential of the diffeomophism relating the coordinate systems of $S_0$ and $S'$. That relation of course exists, but is not deployed in the formalism. Only a constrained form of the differentials are used. The constrained form of the differential evidently is not required to satisfy the full relation (\ref{completediff}) and commutation of the second derivatives (\ref{commutation}). \\
Finally, the gravitational field is claimed to affect the material constituents so that the ruler is \emph{isotropically} scaled, $(g_{ii})^{1/2} = \Phi$. The assumption of isotropy is not stringent, but we have shown that in this simpler model ---in terms of symmetry--- such a constraint on the scaling is in  concordance with 1-PN mechanics in GRT. Moreover in that restricted case we have till that order successfully used $ \Phi = (g_{00})^{-1/2}$. It is however possible to generalize the GMLT,  eq. (\ref{simpleGMLT}), with higher order distinct effects on space and time:
\begin{eqnarray}
\delta t =  \Omega^{-1} \delta \tau' &,&  \delta \mathbf{x} = \Phi \delta \mbox{\boldmath$\xi$}' \label{nonsimpleGMLT}
\end{eqnarray}
This type of generalization ---or anisotropic variants--- could be invoked to fit higher order PN terms.

\end{document}